\RequirePackage[mathlines]{lineno}
\documentclass[aps,titlepage,12pt]{revtex4}
\usepackage{amsfonts}
\usepackage{amsmath}
\usepackage{graphicx}
\usepackage{subfigure}
\usepackage{dcolumn}
\usepackage{bm}
\usepackage{overpic}
\usepackage{booktabs}
\usepackage{color}
\usepackage{multirow}
\usepackage{cases}
\usepackage{amssymb}
\begin{document}
\title{Enhancing speed of pinning synchronizability: low-degree nodes with high feedback gains}
\author{Ming-Yang Zhou$^{1,3}$, Zhao Zhuo$^1$, Hao Liao$^{2,3}$\footnote{jamesliao520@gmail.com}, Zhong-Qian Fu$^1$ and Shi-Min Cai$^4$ \footnote{shimin.cai81@gmail.com}}

\affiliation{1, Department of Electronic Science and Technology, University of Science and Technology of China, Hefei 230027, P. R. China.\\
2, Guangdong Province Key Laboratory of Popular High Performance Computers, College of Computer Science and Software Engineering, Shenzhen University, Nanhai Avenue 3688, Shenzhen 518060, P. R. China.\\
3, Physics Department, University of Fribourg, Chemin du Mus$\acute{e}$e 3, 1700 Fribourg Switzerland. \\
4, Web Sciences Center, School of Computer Science and Engineering, University of Electronic Science and Technology of China, Chengdu 611731, P. R. China.}


\begin{abstract}\noindent
Controlling complex networks is of paramount importance in science and engineering. Despite recent efforts to improve controllability
and synchronous strength, little attention has been paid to the speed of pinning
synchronizability (rate of convergence in pinning control) and the corresponding pinning node selection. To address this issue, we propose
a hypothesis to restrict the control cost, then build a linear matrix inequality related to the speed of pinning controllability.
By solving the inequality, we obtain both the speed of pinning controllability and optimal control strength (feedback gains in pinning control) for all nodes.
Interestingly, some low-degree nodes are able to achieve large feedback gains, which suggests that they have high influence on controlling system. In addition, when choosing nodes with high
feedback gains as pinning nodes, the controlling speed of real systems is remarkably
enhanced compared to that of traditional large-degree and large-betweenness selections. Thus, the proposed
approach provides a novel way to investigate the speed of pinning controllability
and can evoke other effective heuristic pinning node selections for large-scale systems.
\end{abstract}
\maketitle

\textbf{\large Introduction}\\

Swarm, transportation, and many other natural and man-made systems can be represented by networks, in which the
nodes correspond to the agents of systems and the edges describe the relations between the agents
\cite{barabasi2011network, barabasi2012network, newman2010networks, zhang2007heat}.
Some special parts of agents (or units) in these systems adjust their behaviors
on the basis of their surroundings (e.g. location, temperature, taste), while the other parts of agents move according to their neighbors \cite{wang2010control, yan2012controlling}. Consequently, these special agents could influence the dynamics through connectivity of the system and steer the system to a desired state (e.g. location, coordinate).
For example, scouts guide a swarm to fly to a new nest site: When a swarm flies to a new nest site, only 5\% scouts know the right direction and other common bees fly according to their neighbors. In most cases, the swarm reaches the new home \cite{wang2010control}.
Since network connectivity has profound influence on dynamic behaviors (e.g., synchronization, consensus),
analyzing their interplay has attracted scientists from various fields, such as physics, computer science, sociology and others \cite{chavez2005synchronization,arenas2008synchronization,kenett2014network,medo2011temporal,medo2014statistical,yeung2012competition}.
In control problem, controllability of a network relates to both the network connections and the set of driver nodes \cite{liu2011controllability,roukny2013default,karimi2013threshold,jalili2015optimal}. Thus, utilizing network connections to select appropriate driver nodes is a frontier area of research in complex networks\cite{wang2010control,yang2012node,gao2014targeted}.

Beginning with the network perspective, there are two main approaches to assess controllability: algebraic control and structural control. The algebraic approach is the most general and is typically used to investigate control problems \cite{sorrentino2007controllability}, while structural controllability is a simplified analysis that is appropriate for large-scale networks \cite{liu2011controllability,yan2012controlling}. Control problems in general are about steering each node to any arbitrary states. However, in some large-scale network scenarios, we are concerned with the network consensus that is a sub-case of the general network control domain. Pinning control, therefore, which focuses on controlling all the nodes into the same time evolution, has attracted much attention recently \cite{sorrentino2007controllability,yu2013synchronization,yu2013step}.
In the past few years, Wang \emph{et al.} studied the pinning control in scale-free model
networks and showed that selection of high-degree nodes performed better than random selection \cite{wang2002pinning,wang2010control}. But high-degree
selection performs bad in real networks due to the clustering and hierarchical structures in the diffusion process \cite{kitsak2010identification}. Further, Jalili \emph{et al.} explored optimal pinning control in scale-free model network and found pinning nodes had high centrality in scale-free model networks \cite{jalili2015optimal,jalili2013enhancing}.
Liu \emph{et al.} explored the structural controllability of real networks by measuring the minimum number of driver nodes and
found that the number of driver nodes required for full control was determined by the degree distribution \cite{liu2011controllability,cowan2012nodal,shields1975structural}.
Tang \emph{et al.} identified controlling nodes in neuronal networks and found a transition in choosing driver nodes from high-degree
to low-degree nodes \cite{tang2012identifying}. Other issues such as energy cost of controlling a network and the performance
of a single controller have also been investigated \cite{yan2012controlling,olfati2006flocking,yu2013step}.

Our study takes a different, but complementary approach to controllability problem than previous researches that only concerns
whether a network could be controlled and how to improve the range of coupling strength \cite{wang2002pinning,sorrentino2007controllability, wu2008relationship,lu2013theory,yuan2013exact,jia2013control,liu2012control,wang2012optimizing}.
We focus on enhancing the speed of pinning controllability and determining corresponding pinning nodes, where speed of pinning controllability represents the rate of convergence in the control paths and is a more interesting problem in engineering. To enhance speed of pinning controllability, an effective way is controlling every node directly, yet it is only appropriate
for small-scale networks \cite{bubnicki2005modern}. Inspired by some natural flocking phenomena \cite{wang2010control}, we only need to drive a small fraction of nodes to enhance the speed. To address
this key issue, we investigate the speed of pinning controllability and the optimal feedback gains of nodes under restricted control cost in the paper. Our main results show that some low-degree nodes obtain high feedback gains. Further, by choosing pinning nodes with high feedback gains, the speed of pinning controllability is enhanced remarkably
compared with that of traditional methods which select pinning nodes based on their degree or betweenness.
Our method offers an opportunity to investigate the speed of pinning controllability and characteristics of
efficient sets of pinning nodes, which may inspire other better fast heuristic approaches for large-scale complex networks in the future.

\noindent \\ \textbf{\large Results}

In this section, we firstly describe the metrics for the speed of pinning controllability. Next, we introduce the restriction hypothesis of control efficiency and the approach to solve the problem.
At last, to illustrate the validity of our method, the proposed approach is applied to both artificial model and real-world networks.
The results not only demonstrate the effectiveness of the proposed approach but also uncover the characteristics of pinning nodes. Table \ref{table:variable} gives a list of symbols used in this paper.

\begin{table}
\caption{\label{table:variable}Variable notations in the paper.}
\begin{ruledtabular}
\begin{tabular}{c|lp{5in}}
Variable&Description\\\hline
$N$ & Network size\\
$\textbf{x}_i$ & The state variable of node $i$\\
{{A}} & Adjacency matrix of a network\\
$a_{ij}$ & The element of matrix {{A}}\\
$l$ & Size of pinning nodes \\
$\delta$ & Fraction of pinning nodes with $\delta=\frac{l}{N}$\\
$\Gamma$ & Coupling matrix\\
$c$ & Coupling strength\\
$f(x)$& Intrinsic dynamics of a node\\
$d_i$ & Control strength (feedback gain) of node $i$\\
${{D}}$ & Feedback matrix with element $d_{ii}$ being the feedback gain of node $i$\\
${B}$ & {{A}}-{{D}}\\
$\rho$ & A constant related to a dynamical system\\
$\lambda_i({B})$ & $i_{th}$ largest eigenvalue of matrix ${B}$ with $\lambda_N<...<\lambda_2<\lambda_1$\\
$\eta_k$ & Variables related to the states of a network\\
$\partial f(x)$ &  The Jacobian of $f$ on $x$\\
$\lambda_{11}(x(\lambda_k))$ & The largest eigenvalue of  $x(\lambda_k)$ with $\lambda_k$ being the eigenvalues of ${{B}}$\\
$w_i$ & Importance of node $i$\\
$E_i$ & Control efficiency of node $i$\\
$\textbf{w}$ & Vector of nodes' importance, $\textbf{w}=\{w_1^\frac{\alpha}{2},w_2^\frac{\alpha}{2},...,w_n^\frac{\alpha}{2}\}^{'}_{N\times 1}$ with $\alpha$ a tunable parameter
\end{tabular}
\end{ruledtabular}
\end{table}

\textbf{Speed metrics of pinning controllability.}
We start by introducing the stable condition and the metrics to evaluate speed of pinning controllability. To analyze pinning controllability of complex networks, we denote that a connected network
consists of $N$ identical linearly and diffusively coupled nodes, with each node being a $n$-dimensional system.
The state equations of a network are as follows \cite{wang2002pinning}:
\begin{linenomath}
\begin{equation}
\label{equation:self}
\dot{\textbf{x}}_i=f({{\mathbf{x}}_{i}})+c\sum\limits_{j=1}^{N}{{{a}_{ij}}\Gamma {{\mathbf{x}}_{j}}},\quad i=1,2,...,N,
\end{equation}
\end{linenomath}
where $\textbf{x}_i=(x_{i1},x_{i2},...,x_{in})^{'}$, $c$, $\Gamma\in R^{n\times n}$ and $a_{ij}$ are the state variables of node $i$, the coupling strength($c>0$), a matrix linking coupled variables $\Gamma> 0$ and the elements of the adjacency matrix ${A}$, respectively. For the matrix ${A}$, if there is an edge between node $i$ and $j$ ($i\neq j$), then $a_{ij}=a_{ji}=1$; $a_{ij}=a_{ji}=0$ otherwise. Elements $a_{ii}$ of the diagonal are $a_{ii}=-k_i$ with $k_i$ the degree of node $i$.

In Eq. \ref{equation:self}, states of nodes rely on both the intrinsic dynamics of nodes and connectivity of neighbors.
Suppose that we want to stabilize the network on a homogeneous stationary equilibrium \cite{wang2002pinning,wang2010control},
\begin{linenomath}
\begin{equation}
\label{equation:homogeneous}
\begin{cases}
\textbf{x}_1=\textbf{x}_2=...=\textbf{x}_n=\bar{\textbf{x}},\\
f(\bar{\textbf{x}})=0.
\end{cases}
\end{equation}
\end{linenomath}

To achieve the homogeneous state, a typical approach is to select a small fraction $\delta$ $(0<\delta<1)$ of nodes as pinning nodes (denoted by $i_1$, $i_2$, ..., $i_l$)
and apply local linear feedback injections to these pinning nodes.
State equations of pinning nodes are modified as
\begin{linenomath}
\begin{equation}
\label{equation:selfcontrol}
\dot{\textbf{x}}_{i_k}=f({{\textbf{x}}_{i_k}})+c\sum\limits_{j=1}^{N}{{{a}_{i_kj}}\Gamma {{\textbf{x}}_{j}}}-cd_{i_k}\Gamma(\textbf{x}_{i_k}-\bar{\textbf{x}}),\\ k=1,...,l,
\end{equation}
\end{linenomath}
where $d_{i_k}$ is the control strength (Control strength refers to feedback gain in pinning control) of node $i_k$ ($d_{i_k}>0$). Note that Equation \ref{equation:selfcontrol} is reduced to Eq. \ref{equation:self} if all feedback gains equal 0 ($d_{i_k}=0$, for $\forall k, k=1,2,...,l$).

To investigate the speed of pinning controllability, a necessary prerequisite is that the network is stable.
A network can be stabilized onto $\bar{\textbf{x}}$ if the following condition are met\cite{wang2002pinning,sorrentino2007controllability,jalili2015optimal}:
\begin{linenomath}
\begin{equation}
\label{equation:restriction3}
\begin{split}
&c\geq c_{min}=|\frac{\rho}{\lambda_1({B})}|,\\
&or\\
&\sigma_1<c\lambda_N({B})<c\lambda_1({B})<\sigma_2,(\frac{\lambda_N}{\lambda_1}<\frac{\sigma_1}{\sigma_2}),
\end{split}
\end{equation}
\end{linenomath}
where $\rho$, $\sigma_1$ and $\sigma_2$ are constants related to the nodal dynamics of the network \cite{cowan2012nodal}
and $\lambda_i({B})$ are the eigenvalues of matrix ${B}$ that is defined as
\begin{linenomath}
\begin{equation}
\label{equation:matrixB}
{B=A-D}, \quad {D}=diag\{d_1,d_2,..,d_N\}.
\end{equation}
\end{linenomath}

Under the constraints of Eq. \ref{equation:homogeneous} and $\Gamma>0$, the stable condition used in the paper is $c\geq c_{min}=|\frac{\rho}{\lambda_1({B})}|$ \cite{wang2010control,wang2002pinning}. For some other nodal dynamics, the stable condition may be $\sigma_1<c\lambda_N({B})<c\lambda_1({B})<\sigma_2$ that is usually simplified as $\frac{\lambda_N}{\lambda_1}<\frac{\sigma_1}{\sigma_2}$ \cite{sorrentino2007controllability,jalili2015optimal,chavez2005synchronization}. For more information about the stable condition, please refer to the supplementary or Ref. \cite{jalili2015optimal,wang2010control,wu2008relationship}.

Since the stability of Eq. \ref{equation:selfcontrol} is equivalent to $n$ independent equations \cite{wang2002pinning,jalili2015optimal}:
\begin{linenomath}
 \begin{equation}
\label{equation:transferfunction}
\dot{\eta}_k=[\partial f(\bar{\textbf{x}})+c\lambda_k\Gamma]\eta_k,
\end{equation}
\end{linenomath}
where $\eta_k$ are variables related to states of nodes. $\partial f(\bar{\textbf{x}})$ is the Jacobian of $f$ on $\bar{\textbf{x}}$. Suppose that the system is stable, the speed of pinning controllability is determined by the largest eigenvalue $\lambda_{11}$ of $[\partial f(\bar{\textbf{x}})+c\lambda_k\Gamma]$ that takes over all $\lambda_k (k=1,2,...,N)$:
\begin{linenomath}
 \begin{equation}
\label{equation:speed}
v=max\{\lambda_{11}([\partial f(\bar{\textbf{x}})+c\lambda_k\Gamma]), k=1,2,...N\},\quad v<0,
\end{equation}
\end{linenomath}
where $\partial f(\bar{\textbf{x}})$ is the the Jacobian of $f$ on $\bar{\textbf{x}}$ and $\lambda_k$ $(0>\lambda_1>\lambda_2>...>\lambda_N)$ are the eigenvalues of ${B}$.
Note that, unlike previous researches about expanding interval of coupling strength in Eq. \ref{equation:restriction3} that only requires $v<0$. Equation \ref{equation:speed} characterizes the rate of convergence that relate to all eigenvalues $\lambda_i$,$i=1,2,...N$. Larger $|v|$ represents higher rate of convergence of the system. Therefore, enhancing the speed is equivalent to increasing $|v|$.

Further, since $\lambda_{11}([\partial f(\bar{\textbf{x}})+c\lambda_k\Gamma])=max\{\frac{\mathbf{y}^T\cdot([\partial f(\bar{{x}})+c\lambda_k\Gamma])\cdot \mathbf{y}}{\mathbf{y}^T \cdot \mathbf{y}}, \forall \mathbf{y}\in R^{n\times 1}, \mathbf{y}\neq 0\}$, under the constraints of Eq. \ref{equation:homogeneous} and $\Gamma> 0$,
for any two eigenvalues $\lambda_i$ and $\lambda_j$ ($\lambda_i<\lambda_j<0$),
\begin{linenomath}
 \begin{equation}
 \begin{split}
 {{\lambda }_{11}}([\partial f(\overline{\mathbf{x}})+c{{\lambda }_{i}}\Gamma ])&=max\{\frac{{\mathbf{y}^{T}}\cdot ([\partial f(\overline{x})+c{{\lambda }_{i}}\Gamma ])\cdot \mathbf{y}}{{\mathbf{y}^{T}}\cdot \mathbf{y}}\} \\
 & =max\{\frac{{\mathbf{y}^{T}}\cdot ([\partial f(\overline{x})+c{{\lambda }_{j}}\Gamma ])\cdot \mathbf{y}}{{\mathbf{y}^{T}}\cdot \mathbf{y}}+\frac{c({{\lambda }_{i}}-{{\lambda }_{j}})\cdot {\mathbf{y}^{T}}\cdot \Gamma \cdot \mathbf{y}}{{\mathbf{y}^{T}}\cdot \mathbf{y}}\} \\
 & <max\{\frac{{\mathbf{y}^{T}}\cdot ([\partial f(\overline{x})+c{{\lambda }_{j}}\Gamma ])\cdot \mathbf{y}}{{\mathbf{y}^{T}}\cdot \mathbf{y}}\} \\
 & ={{\lambda }_{11}}([\partial f(\overline{\mathbf{x}})+c{{\lambda }_{j}}\Gamma ]). \\
  \end{split}
 \end{equation}
\end{linenomath}

 Thus, $\lambda_{11}([\partial f(\bar{\textbf{x}})+c\lambda_1\Gamma])>\lambda_{11}([\partial f(\bar{\textbf{x}})+c\lambda_i\Gamma]),(i=2,3,...,N)$. Equation \ref{equation:speed} can be simplified as
\begin{linenomath}
 \begin{equation}
\label{equation:speed2}
v=\lambda_{11}([\partial f(\bar{\textbf{x}})+c\lambda_1\Gamma]).
\end{equation}
\end{linenomath}

Since $\lambda_1<0$ and $v<0$, lower $\lambda_1$ represents higher absolute $|v|$ and higher rate of convergence in the control processes. $\lambda_1({B})$ determines the speed $\lambda_{11}([\partial f(\bar{\textbf{x}})+c\lambda_1\Gamma])$.  Thus, $\lambda_1({B})$
is positive correlated with the speed $\lambda_{11}([\partial f(\bar{\textbf{x}})+c\lambda_1\Gamma])$. Therefore, lower $\lambda_1({B})$ is better.

In some master-slave natural and man-made systems, the states of pinning nodes are fixed to the  homogeneous state, which could be represented by applying infinite feedback gains to the pinning nodes in mathematics \cite{guyton1990surprising,wang2002pinning}. Therefore, we apply infinite feedback gains to the pinning nodes ($d_i\rightarrow\infty$ for these nodes) and no feedback gains to the other nodes ($d_i=0$ for other nodes) \cite{wang2010control,wang2002pinning}. Then, the eigenvalue $\lambda_1({B})$ equals to $\lambda_1(\bar{{A}})$ \cite{wang2002pinning}:
\begin{linenomath}
 \begin{equation}
\label{equation:matrixB}
\underset{\begin{smallmatrix}
 {{d}_{{{i}_{k}}}}\to \infty , \\
 \forall k,k=1,...,l
\end{smallmatrix}}{\mathop{\lim }}\,{{\lambda }_{1}}({B})={{\lambda }_{1}}(\bar{{A}}),
\end{equation}
\end{linenomath}
where $\bar{{A}}$ is obtained by removing the $i_1-th$, $i_2-th$,...,$i_l-th$ row and $i_1-th$, $i_2-th$,...,$i_l-th$ column of ${A} $ \cite{wang2002pinning}, and $\lambda_1{(\bar{{A}})}$ is the largest eigenvalue of matrix $\bar{{A}}$. In the following, based on the positive correlation between $\lambda_1{(\bar{{A}})}$ and the speed $v$, we thus use  $\lambda_1{(\bar{{A}})}$ as the metric to evaluate the speed of controllability for a specific set of pinning nodes.

Since the pinning node selection plays an important role in the speed of pinning controllability, to enhance the speed of pinning controllability, the key issue is how to select an appropriate set of pinning nodes. However, for the fixed size $l$ of pinning nodes, it is computationally prohibitive to select $l$ pinning nodes from a network
of size $N$ because there are $C^{l}_{N}$ cases of different combinations. A feasible solution is to propose efficient heuristic
approach that approximately matches optimal selection. Traditional approaches usually select pinning nodes according to
nodes' importance, such as degree and betweenness. However, though a single important node has a great influence on the dynamics,
multiple important nodes may performs bad due to overlapping influences of these nodes. Thus, adding extra nodes with high importance does not benefit the speed of pinning controllability effectively. Consequently, to design heuristic approaches, we need to explore the characteristics of effective multiple pinning nodes. Thus, we propose a restriction about control efficiency and utilize \emph{linear Matrix Inequality} (LMI) method to solve the problem.

\textbf{Restriction of control efficiency and solution of optimal feedback gains.}
In this section, an approach is proposed to calculate the optimal $\lambda_1({B})$ and feedback gain $d_i$ for each node.
Our approach firstly build the relation between feedback gain $d_i$ and importance (e.g., degree and betweenness) of node $i$ with control efficiency. Based on that, an inequality is constructed and solved to obtain optimal $\lambda_1(B)$ and the corresponding feedback gains for all nodes.

The first step is to give the restriction about control efficiency. To control a network, it is usually efficient to steer high important nodes \cite{wang2002pinning,jalili2015optimal}. The importance of nodes plays a significant role in controllability, where importance is usually characterized by degree, betweenness, etc. Besides, control cost of a node is directly related to its feedback gains with positive correlation. Thus, control efficiency $E_i$ of node $i$ is defined as follows:
\begin{linenomath}
\begin{equation}
\label{equation:cost}
E_i=d_i\cdot w_i^\alpha,
\end{equation}
\end{linenomath}
where $w_i$ is the importance of node $i$ and $\alpha$ varies from -1 to 0.
We denote $w_i=k_i$ (degree) in the paper and $w_i=g_i$ (betweenness) in the supplementary, respectively. Lower $E_i$ represent higher efficiency. For the fixed $d_i$, high-important nodes should have high efficiency.


Since important nodes play a key role in the dynamics of networks \cite{wang2006traffic,yan2006efficient,zhou2012traffic},
we propose a hypothesis that a network has limited control efficiency $C$, which follows
\begin{linenomath}
\begin{equation}
\label{equation:sumcost}
E_{sum}=\sum E_i=\sum d_i\cdot w_i^\alpha=C.
\end{equation}
\end{linenomath}

For the fixed $E_{sum}$ and $\alpha$ $(\alpha<0)$,
nodes with large $w_i$ tend to have low $w_i^\alpha$ and large $d_i$.
Thus, high-important nodes have more probability to be chosen as pinning nodes.

Based on the restriction of control efficiency, we then transfer the speed of pinning controllability problem into a LMI problem.
For a given network, the aim is to find an optimal ${D}_{opt}$ which minimizes
the largest eigenvalue $\lambda_1({B})$ of matrix ${B}$:
\begin{linenomath}
\begin{equation}
\label{equation:minlamda}
{D}_{opt}=\{{D}|min\lambda_1({B}), \quad {B=A-D}\},
\end{equation}
\end{linenomath}
where ${D}$ is an unknown diagonal matrix variable in which elements on the diagonal are the feedback gains of the corresponding nodes.

Through some mathematical transition, the speed of pinning controllability and optimal feedback gains are also equivalent to a LMI function in which $\lambda_1({B})_{min}=\lambda_{x,optimal}$:
\begin{linenomath}
\begin{equation}
\label{equation:transfereq}
{B}={A}-{D}<\lambda_x {I},
\end{equation}
\end{linenomath}
where ${I}$ is the identity matrix.
$\lambda_x$ is the unknown variable and the aim is to search optimal ${D}$ that minimizes $\lambda_x$.

If $\alpha=0$, $w_i$ reduces in Eq. \ref{equation:sumcost} and $\sum d_i=C$. The optimal solution for Eq. \ref{equation:transfereq}
is $\lambda_{x,min}=-\frac{C}{N}$ and ${D}=\frac{C}{N}{I}$ at $\alpha=0$, which implies that all nodes obtain identical feedback gains
and the difference of nodes can't be distinguished by feedback gains. For more details, please refer to
Eq. \ref{equation:solveeigvalue1}$-$\ref{equation:solveeigvalue3}.

The analytic solution ${D}_{optimal}$ and $\lambda_x$ are obtained merely at $\alpha=0$.
For $\alpha\neq0$, we get the numerical solution under the restriction Eq. \ref{equation:sumcost}. Equation \ref{equation:sumcost} and \ref{equation:transfereq} construct a standard
LMI problem that can be solved by convex optimization methods \cite{boyd1994linear,tanaka2004fuzzy} and Interior-Point Methods \cite{boyd1994linear}. Through the inequality optimization (Eq. \ref{equation:constraint0} $-$\ref{equation:constraint4}), we obtain optimal feedback gains for each node and the optimal $\lambda_1({B})=\lambda_{x,min}$.

\begin{figure*}[!htp]
\centering
\includegraphics[width=6.5in]{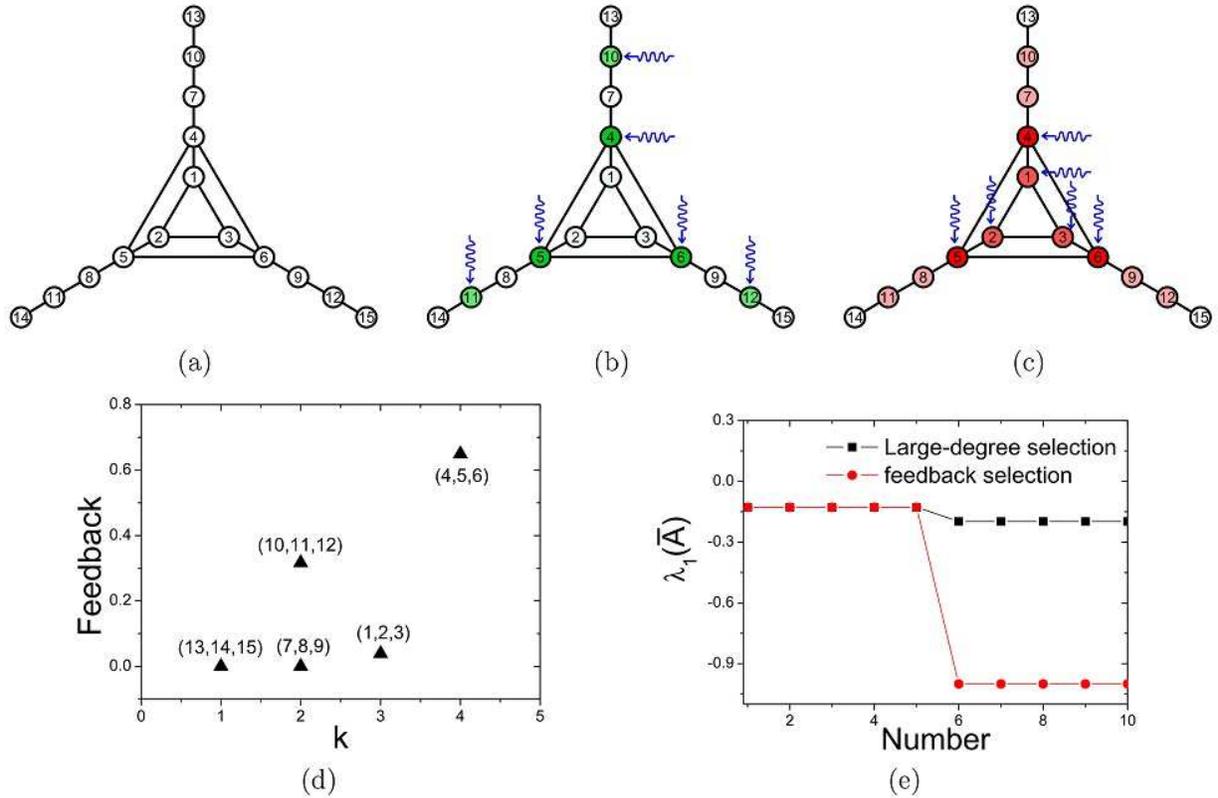}
  \caption{(Color online) Illustration of the optimal feedback gains for an artificial network (The size of pinning nodes is fixed $Number=6$ for sub-figure (b) and (c) ).
  (a) A simple undirected and unweighted network. (b) Six pinning nodes selected according to the feedback gains of nodes (Dark green represents higher feedback gains).
  (c) Six pinning nodes selected according to the degree of nodes ( Dark red represents higher degree).
  (d) The relation between feedback gains and degree for the artificial network. Numbers in the subfigure represent labels of nodes.
  (e) The largest eigenvalue $\lambda_1$ of $\bar{{A}}$ represents the speed of pinning controllability for the network.
  Lower $\lambda_1(\bar{{A}})$ indicates higher speed of pinning controllability and the proposed approach has better performance when $Number\geq6$.}
  \label{fig:illustrationdemo}
\end{figure*}

\textbf{Pinning node selection.} In the selection process, we first calculate the optimal feedback gains for all nodes by LMI method. Furthermore, for the fixed size $l$ of pinning nodes,
nodes with high feedback gains are chosen as pinning nodes. Next, the selected pinning nodes are  injected infinite feedback gains and other nodes obtain none feedback gains. The performance of our approach, which is evaluated by $\lambda_1(\bar{{A}})$, is compared with large-degree selection method.
The proposed approach on a small artificial network is illustrated in Fig. \ref{fig:illustrationdemo}. Figure \ref{fig:illustrationdemo}(d) shows that feedback gains of nodes are obviously different from their degrees
and more interestingly some low-degree nodes (e.g., node 10, 11, and 12) obtain high feedback gains, which suggests that the feedback gains of nodes are determined by both degree and structure of the network. Based on the feedback gains, we then select pinning nodes according to feedback gains, in comparison with traditional high-degree selection (see Fig. \ref{fig:illustrationdemo}(b) and \ref{fig:illustrationdemo}(c)).
Figure \ref{fig:illustrationdemo}(e) shows the speed of pinning controllability as a function of size of pinning nodes, in which the speed is obviously enhanced compared with large-degree selection when the size exceeds $6$ ($Number\geq6$).

\textbf{Results in BA model and real networks.}
The validity of our proposed approach is verified in four undirected and unweighted networks with different backgrounds: a BA model network, a power grid network (PowerGrid), a biological network (PDZBase) and a social network (Jazz).
The BA model network is generated from a small number of connected nodes and every new node links $m$ edges to
$m$ existing nodes with preferential probability \cite{barabasi1999emergence}. The probability that a new node links to node $i$ depends on the degree $k_i$ of node $i$,
such that $\prod(k_i)=\frac{k_i}{\sum_{j\in{N}}{k_j}} $ ($m=3$ in the paper). BA model network has 300 nodes and 893 edges. PowerGrid is the power grid of the Western States of the United States of America \cite{watts1998collective}. In order to reduce computation complexity, we extract 3-core of PowerGrid that only reserves nodes with degree larger than 3. The extracted subnetwork has 116 nodes and 217 edges and keeps similar structures with primitive network due to self-similarity properties of complex networks \cite{song2005self,ravasz2002hierarchical,ravasz2003hierarchical}. PDZBase is a biological network of protein-protein interactions from PDZBase with 161 nodes and 209 edges \cite{beuming2005pdzbase}. Jazz is a cooperation social network with 198 nodes and 2742 edges \cite{gleiser2003community}.

\begin{figure*}[!htp]
\centering
\includegraphics[width=6in]{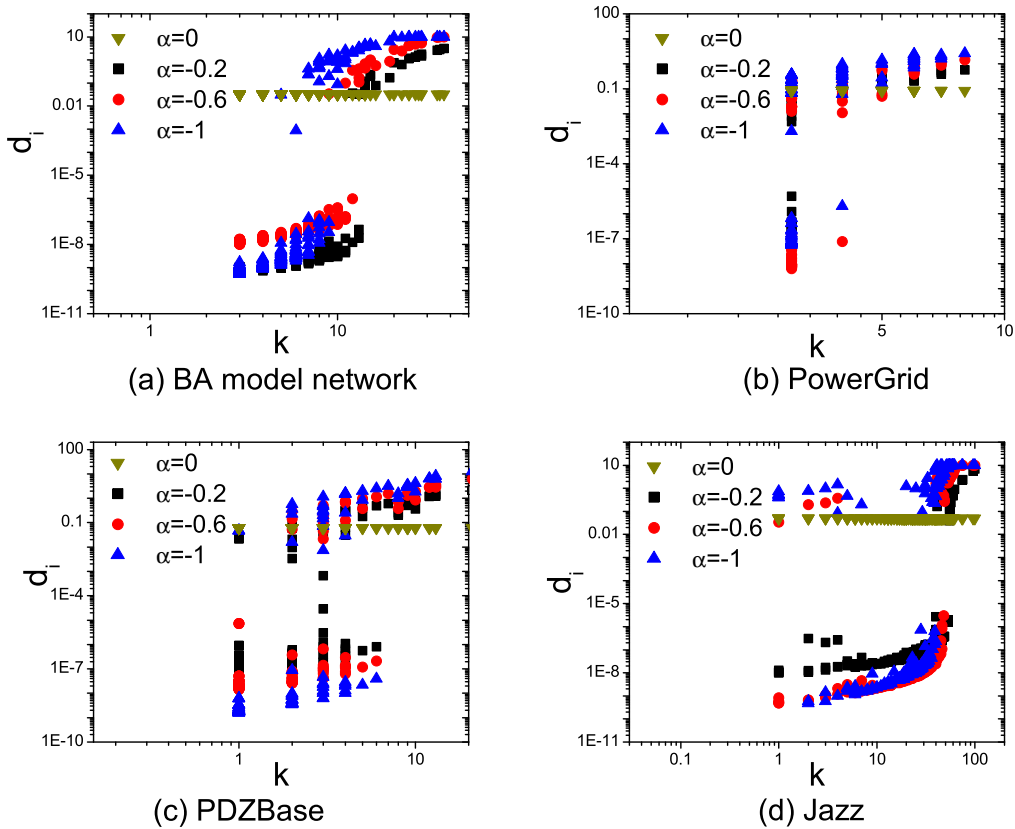}
  \caption{(Color online) The distributions of feedback gains as a function of $k$ for four networks in restriction of $\alpha=0, -0.2, -0.6, -1$ at $C=10$.
  The results are obtained by LMI optimization, and accuracy of $\lambda_x$ is $1\times10^{-6}$ in the optimization process. A positive correlation exists between feedback gain and degree in BA model networks. However in real-world networks, many low-degree nodes have high feedback gains, which suggests that the feedback gains depend on not only their degree but also
the connectivity of networks.}
  \label{fig:feedbackalpha}
\end{figure*}

Given a network, the inequality (Equation \ref{equation:transfereq}) is restricted by both the control efficiency and tunable parameter $\alpha$. We firstly explore the influence of $\alpha$ on the results.  For $C=10$, Figure \ref{fig:feedbackalpha} depicts
the distribution of feedback gains with different $\alpha$.
In Fig. \ref{fig:feedbackalpha}(a), if $\alpha\neq 0$, high-degree nodes tend to get high feedback gains and low-degree ones almost get no feedback gains ($d_i\approx0$), which indicates that the feedback gains of nodes are associated with their degree
and pinning nodes can be selected according to the degree in BA model networks. However, results in real networks are different from BA model network.
Figure \ref{fig:feedbackalpha}(b)$-$(d) show that some nodes with lower degree obtain
largely positive feedback gains ($d_i\gg0$). Moreover,
according to Eq. \ref{equation:sumcost}, it is easy to understand that high-degree
nodes tend to obtain higher feedback gains when $|\alpha|$ increases. Obviously,
it works well in BA model network. Whereas in real networks, as $|\alpha|$ increases,
the feedback gains of some low-degree nodes increase remarkably.
It's because that BA model networks have no community structure, nor hierarchical organization.
These structures in real networks lead to the overlapping influences of pinning nodes.
The new results suggest that better set of pinning nodes should contain both high-degree nodes
and those low-degree ones with high feedback gains.


\begin{figure*}[!htp]
\centering
    \includegraphics[width=6in]{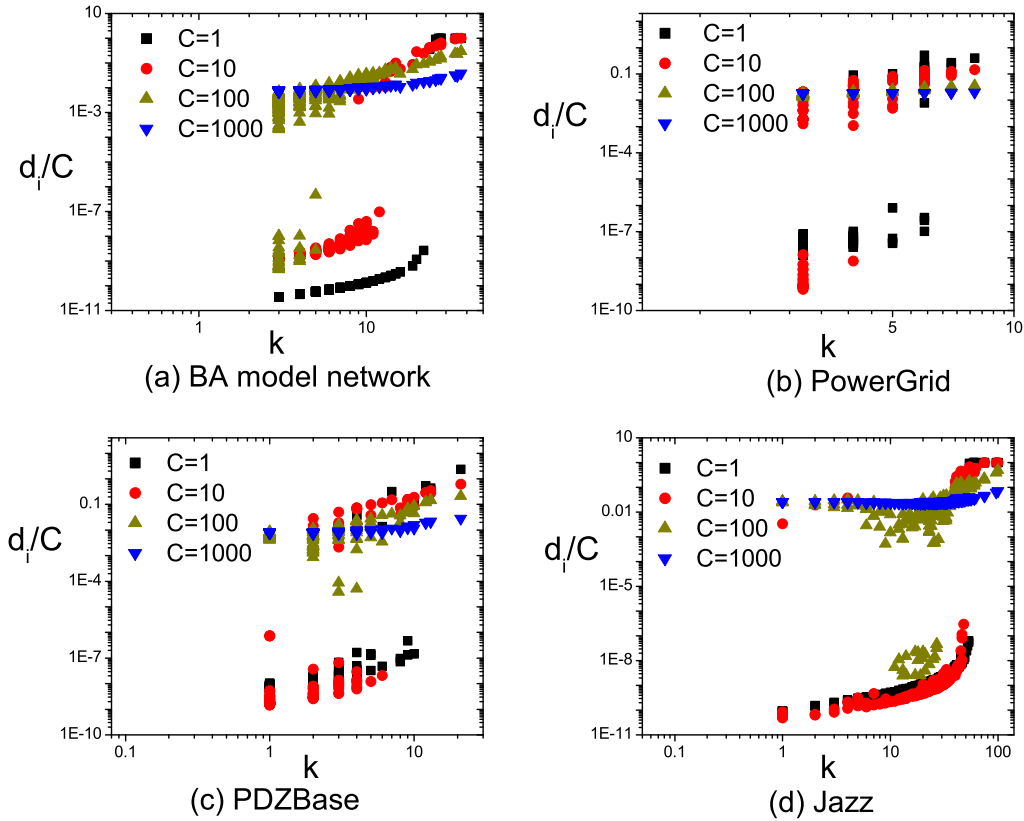}
  \caption{(Color online) The distributions of feedback gains as a function of degree $k$ for four networks in restriction of $C=1,10,100,1000$ at $\alpha=-0.6$.
  The results are got by LMI optimization, and the accuracy of $\lambda_x$ is $1\times10^{-6}$ in the optimization process. The results suggest that
  restriction of control efficiency obviously affects the the distributions of feedback gains, especially when $C$ is small.}
  \label{fig:feedbacksum}
\end{figure*}

The distribution of feedback gains is affected by not only $\alpha$ but also $E_{sum}$. For fixed
$\alpha$, we study the relation between feedback gains and degree of nodes under different $C$ ($E_{sum}=C$). Figure \ref{fig:feedbacksum} depicts the relation
with $C=1, 10, 100, 1000$ and $\alpha=-0.6$, which shows that the gaps of feedback gains become smaller as $C$ increases.
More specifically, for $C=1, 10, 100$, nodes have apparent different feedback gains: some nodes obtain large positive feedback gains, while other nodes get almost none feedback gains.
However, when $C=1000$, nodes have almost the same positive feedback gains. It suggests that if restriction of control efficiency does not exist, controlling nodes directly is more efficient.
When $C$ is small ($C<1$), only a small fraction of nodes could obtain high feedback gains. As $C$ increases, the restriction of control efficiency influences
the differences of feedback gains little by little and more nodes could obtain high feedback gains.

\begin{figure*}[!htp]
\centering
    \includegraphics[width=6in]{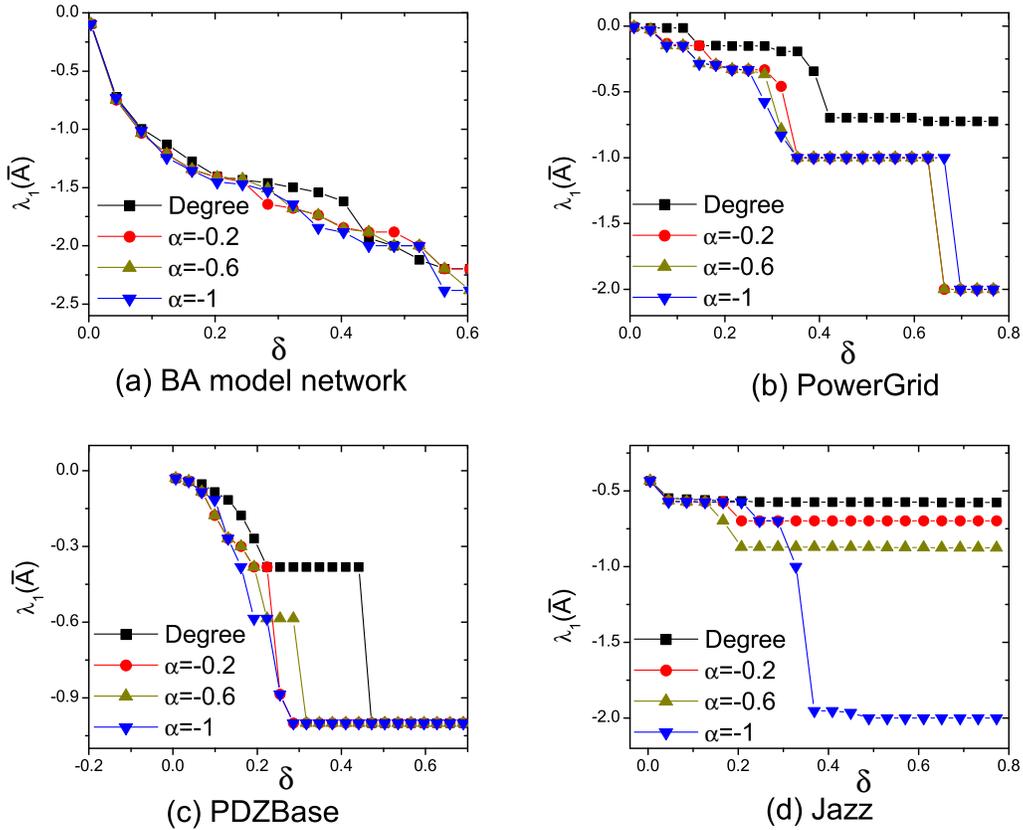}
  \caption{(Color online) The largest eigenvalue $\lambda_1(\bar{{A}})$ as a function of $\delta$ and $\alpha$ for four networks at $C=10$.
   Note that, in large-degree pinning control, pinning nodes are obtained by selecting the largest $\lfloor N\delta\rfloor$ degree nodes.
   The results show that the proposed approach can efficiently enhance the speed of pinning controllability.}
  \label{fig:eigenvaluealpha}
\end{figure*}

To meet real-world conditions, nodes with highest feedback gains are selected as pinning nodes and they are
applied into infinite feedback gains ($l=\lfloor N\delta\rfloor$). Figure \ref{fig:eigenvaluealpha} shows $\lambda_1(\bar{{A}})$
as a function of $\delta$ and $\alpha$ with $C=10$. Comparing method that selects pinning nodes by degree (large-degree selection), our approach has much better performance in real-world networks. Whereas in BA model network, the feedback gain and degree have a high positive correlation. Figure \ref{fig:eigenvaluealpha}(a) shows that they have similar performance. Different from BA model networks, real networks have hierarchical and community structures that
results in overlapping influences \cite{kitsak2010identification}. So large-degree selection has poor performance in real networks.
Our approach overcomes this problem and some low degree periphery nodes obtain high feedback gains. These low degree nodes with high feedback gains
can also enhance the speed of controllability.

Meanwhile, we also test  $\lambda_1( \bar{{A}})$ under different $C$.
The proposed approach has similar results with large-degree selection in BA model network. However,
it performs better than large-degree pinning control in real networks, which is due to the different topology between BA model and real-world networks. Since the result is similar in Fig. \ref{fig:eigenvaluealpha}, more details of different $C$ are shown in the supplementary Fig. S1.

\textbf{Characteristics of pinning nodes.}
Extracting characteristics of effective pinning nodes is interesting when designing fast heuristical approaches. In this section, we mainly investigate two features of effective pinning nodes: the average distance between pinning nodes and average shortest paths from a common node to its nearest pinning node. The results show that increasing
the sparsity between pinning nodes could enhance the speed of pinning controllability.

The average distance $\bar{L}$ between pinning nodes could describe the sparsity of pinning nodes, which follows
\begin{linenomath}
\begin{equation}
\label{equation:averagedistance}
\bar{L}=\frac{1}{|N_d|(|N_d|-1)}\sum_{i,j\in N_d} l_{ij},
\end{equation}
\end{linenomath}
where $N_d$ represents the set of pinning nodes, $l_{ij}$ is the shortest distance from pinning node $i$ to $j$.
Higher $\bar{L}$ indicates sparser pinning nodes.

Another metric to estimate the sparsity is the average of shortest distances from a common node to its nearest pinning node:
\begin{linenomath}
\begin{equation}
\label{equation:averageleastdistance}
\bar{L}_{min}=\frac{1}{N-|N_d|}\sum_{i \notin N_d} min_{j\in N_d}\{l_{ij}\},
\end{equation}
\end{linenomath}
where $min_{j\in N_d}\{l_{ij}\}$ is the shortest distance from a common node $i$ to the set of pinning nodes.

\begin{figure*}[!htp]
\centering
    \includegraphics[width=6in]{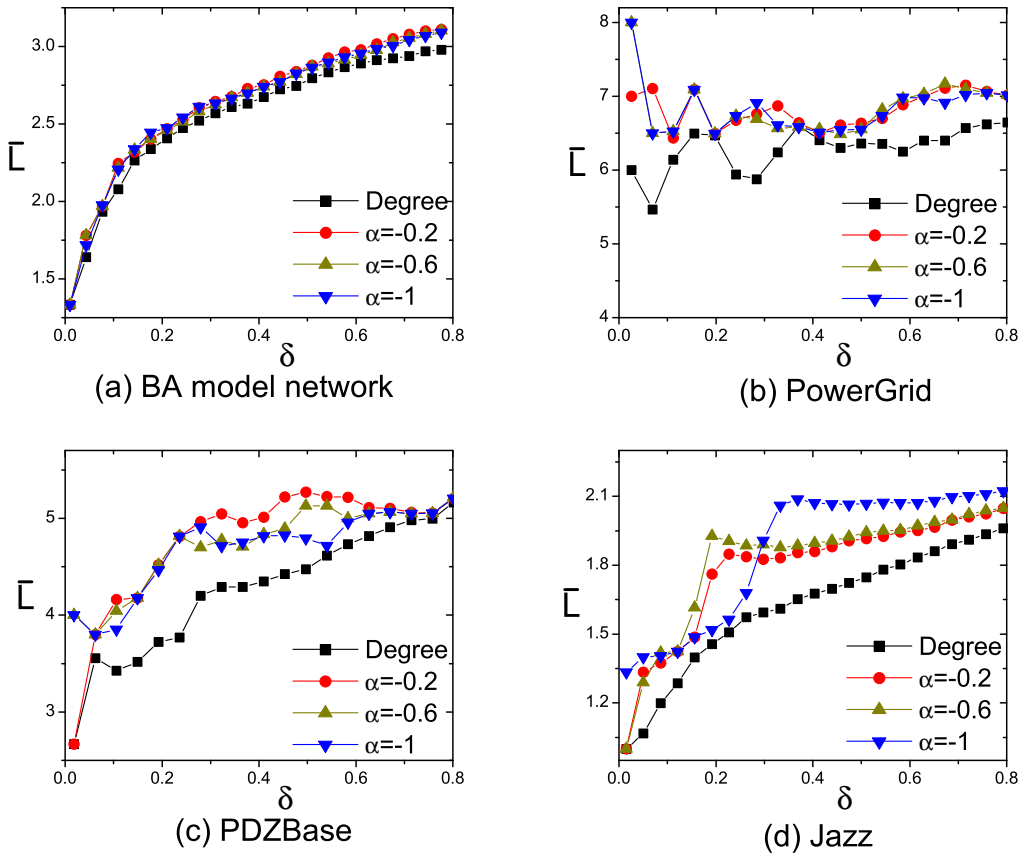}
  \caption{(Color online) The average distance $\bar{L}$ as a function of $\delta$ and $\alpha$ for four networks at $C=10$.
  Note that, large-degree pinning (\emph{Degree}) control where the pinning nodes are got
   by selecting the largest $\lfloor N\delta\rfloor$ degree nodes.}
  \label{fig:averagedistancealpha}
\end{figure*}

Figure \ref{fig:averagedistancealpha} shows $\bar{L}$ of four networks at $C=10$.
In BA model networks (see Fig. \ref{fig:averagedistancealpha}(a)), the proposed approach has almost the same performance with
large-degree selection. The reason is that large degree nodes have large feedback gains in BA model network and the selected pinning nodes are also high degree nodes.
So they have similar results for arbitrary $\alpha(\alpha<0)$ in BA model networks. However, in real networks, the proposed approach selects sparser
pinning nodes than those of large-degree selection. Some periphery low-degree nodes obtain large feedback gains. Hence the sparsity is enhanced.
Figure \ref{fig:averagedistancealpha}(b)$-$(d) show that the sparsity of pinning nodes first increases dramatically, then keeps stable or changes slightly.
By synthesizing Fig. \ref{fig:feedbackalpha} and \ref{fig:averagedistancealpha}, we find that the proposed approach first selects large-degree nodes,
and then selects some lower-degree nodes. The low-degree pinning nodes increase the sparsity. Further, the influence of different $C$ on the results are similar to that in Fig. \ref{fig:averagedistancealpha}. Details about the influence of $C$ are shown in the supplementary Fig. S2.

\begin{figure*}[!htp]
\centering
    \includegraphics[width=6in]{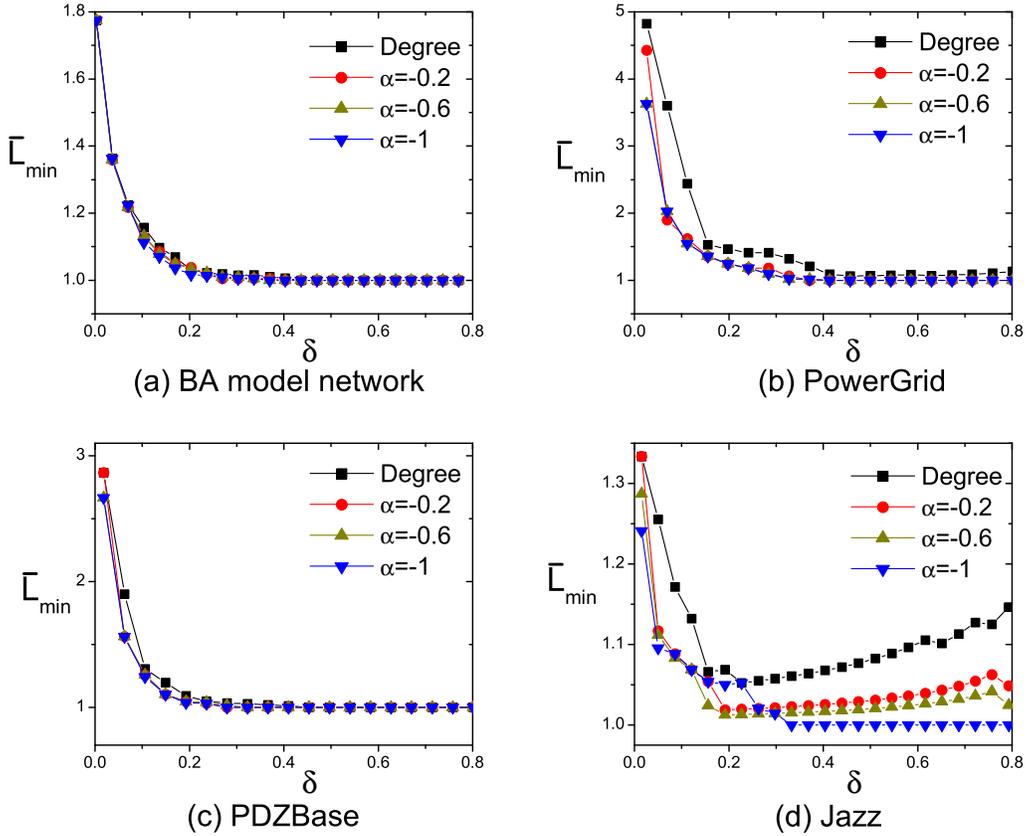}
  \caption{(Color online) The average shortest distance $\bar{L}_{min}$ as a function of $\delta$ and $\alpha$ for four network at $C=10$.
  Note that, large-degree pinning control (\emph{Degree}) where the pinning nodes are got
   by selecting the largest $\lfloor N\delta\rfloor$ degree nodes.}
  \label{fig:leastdistancealpha}
\end{figure*}

Besides the average distance $\bar{L}$, Figure \ref{fig:leastdistancealpha} shows $\bar{L}_{min}$ of four networks at $C=10$.
Our approach and large-degree selection have similar performances in BA model network (see Fig. \ref{fig:leastdistancealpha}(a)).
Because of the positive correlation between feedback gains and degree in Fig. \ref{fig:feedbackalpha}(a), pinning nodes chosen by both approaches are the same.
Apart from BA model network, both methods have similar results in PowerGrid and PDZBase networks except $\delta<0.4$,
which is due to the restricted size of networks. When $\delta<0.4$, our proposed approach selects sparse pinning nodes, leading to a little lower $\bar{L}_{min}$. But as the size of pinning nodes increases, some high-degree nodes are selected, leading to that distances from pinning nodes to the other common nodes are 1. Thus, the differences can't be observed
in the two networks when $\delta>0.4$. However, in jazz network, the proposed approach has lower $\bar{L}_{min}$, which suggests that distance from common nodes to pinning nodes
is reduced. Since lower distance benefits the spreading of control signals, the speed of pinning controllability is enhanced.
Except the influences of $\alpha$, we also explore the influence of $C$ in the supplementary Fig. S3. The results are similar to Fig. \ref{fig:leastdistancealpha}.

\noindent \\ \textbf{\large Discussion}\\

In summary, we systematically study the relations between the speed of pinning controllability and pinning node selection.
Based on the relation between feedback gains and the importance of nodes, we propose a restriction to limit the efficiency of networks.
Then a LMI function is constructed (Eq. \ref{equation:constraint0}--\ref{equation:constraint5}), from which we utilize convex optimization to solve the speed boundary of pinning controllability
and the optimal feedback gains for each nodes. Next, to meet the real-world conditions, we propose a new method to select a small proportion of pinning nodes with high feedback gains
and apply infinite feedback gains to these nodes. The proposed approach achieves remarkable improvements in the speed of pinning controllability for real networks compared to traditional large-degree and and large-betweenness selections.
The results suggest that optimal selection of pinning nodes should contain nodes with both high and low degree. Moreover, unlike previous investigations that only focused on one optimal controller \cite{yu2013step},
we study the characteristics of optimal feedback gains and near-optimal set of multiple pinning nodes.

Though the proposed approach investigates the problem in undirected and unweighted networks, it could also be extended to directed and weighted networks with minor modification.
The presented results have many potential applications in the future. Characteristics of effective pinning nodes could inspire fast heuristic algorithms to choose pinning nodes for large-scale
complex networks in the future. Besides, our method provides a step forward from the current research on controllability
toward enhancing the speed of pinning controllability for complex networks.

\noindent \\ \textbf{\large Methods}\\
\\
\textbf{LMI problems related to speed of pinning controllability.} The speed of pinning controllability is evaluated by $\lambda_1({B})$ and the aim is to search an appropriate diagonal matrix ${D}$ that minimizes $\lambda_1({B})$.
The investigation about speed of pinning controllability and Equation \ref{equation:transfereq} are also equivalent to a LMI function:
\begin{linenomath}
\begin{equation}
\label{equation:constraint0}
\emph{min} \quad (\lambda_x),
\end{equation}
\end{linenomath}
which subjects to
\begin{linenomath}
\begin{numcases}
\textbf{{D}}=\begin{bmatrix}
d_1\\
&d_2& &\text{{\huge{0}}}\\
& & \ddots \\
& \text{{\huge{0}}}& &d_{N-1}\\
& & & & d_N
\end{bmatrix}_{N\times N}, \label{equation:constraintdiagonal}\\
{0}<{D},\label{equation:constraint1}\\
{D}<d_{max}{I},\label{equation:constraint2}\\
{\mathbf{w}^{'}D\mathbf{w}}=C,\label{equation:constraint6}\\
{A}-{D}<\lambda_x {I}\label{equation:constraint5},
\end{numcases}
\end{linenomath}
where $d_{max}$ is the upper bound of feedback gains for all nodes. ${I}$ is an identity matrix in which elements on the diagonal are 1, otherwise 0. $\textbf{w}$ is a $n\times1$ column vector relevant to the importance of the whole nodes ($\textbf{w}=\{w_1^\frac{\alpha}{2},w_2^\frac{\alpha}{2},...,w_n^\frac{\alpha}{2}\}^{'}_{N\times 1}$, $w_i=k_i$ in the paper and $w_i=g_i$ in the supplementary, where $g_i$ is the betweenness of node $i$). $C$ represents the sum of $E_i$ ($C>0$) and Equation \ref{equation:constraint6} is equivalent to Eq. \ref{equation:sumcost}. ${A}-{D}<\lambda_x {I}$ means that (${A}-{D}-\lambda_x {I}$) is negative definite. The constraint Eq. \ref{equation:constraint1} and Eq. \ref{equation:constraint2} confirm that the feedback gain of every node ranges from $0$ to $d_{max}$ ($d_{max}=C$ in the paper). $\lambda_x$ is the desired variable and the aim is to search optimal ${D}$ that minimizes $\lambda_x$.

\textbf{Speed boundary of pinning controllability.} Under the restriction of control efficiency, the upper bound of  speed could be obtained from Eq. \ref{equation:transfereq}. According to Eq. \ref{equation:constraint6} and Eq. \ref{equation:constraint5}, the lower bound of $\lambda_x$ could be obtained by modifying Eq. \ref{equation:constraint5} as
\begin{linenomath}
\begin{equation}
\label{equation:solveeigvalue1}
{A}- \lambda_x {I}<{D}.
\end{equation}
\end{linenomath}
Since (${A}- \lambda_x {I}-{D}$) is negative definite, we obtain
\begin{linenomath}
\begin{equation}
\label{equation:solveeigvalue2}
\mathbf{w}^{'}({A}- \lambda_x {I})\mathbf{w}<\mathbf{w}^{'}{D}\mathbf{w}.
\end{equation}
\end{linenomath}
Substituting Eq. \ref{equation:constraint6} into Eq. \ref{equation:solveeigvalue2}, the boundary of $\lambda_x$ follows as
\begin{linenomath}
\begin{equation}
\label{equation:solveeigvalue3}
\lambda_x>\lambda_{x,min}=-\frac{1}{\sum\limits_{i}{k_i^{\alpha}}}(C-\mathbf{w}^{'}A\mathbf{w}).
\end{equation}
\end{linenomath}

The lower bound of $\lambda_x$ is given in Eq. \ref{equation:solveeigvalue3}, from which we can find that the minimum of $\lambda_x$ is proportional to $C$. Note that, if $\alpha=0$, $\textbf{w}=(1,1,...,1)^{'}$. Since $\lambda=0$ is an eigenvalue of ${A}$ and the corresponding eigenvector is $\textbf{w}=(1,1,...,1)^{'}$, $\mathbf{w}^{'}A\mathbf{w}=0$. Thus, $\lambda_{x,min}=-\frac{C}{N}$ when $\alpha=0$. Moreover, if all nodes have identical feedback gains ($d_i=\frac{C}{N}$ and ${D}=\frac{C}{N}{I}$), the lower bound of $\lambda_x$ is $\lambda_{x,min}=-\frac{C}{N}$. So the optimal feedback gains are $d_i=\frac{C}{N}$ and $\lambda_1({B})=\frac{C}{N}$ when $\alpha=0$.

Though the lower bound of $\lambda_x$ is given in Eq. \ref{equation:solveeigvalue3}, it's difficult to get the analytic solution of matrix ${D}$ for arbitrary $\alpha$. It has been proven that only the numberical solution could be obtained under the restrictions in Eq. \ref{equation:constraint0}--\ref{equation:constraint5} due to its complexity \cite{boyd1994linear,tanaka2004fuzzy}. Restrictions of Eq. \ref{equation:constraint0}--\ref{equation:constraint5} constitute a standard linear matrix inequality(LMI) problem that could be solved by convex optimization methods \cite{boyd1994linear,tanaka2004fuzzy,hautus1969controllability}.
The LMI problem in Eq. \ref{equation:constraint0}--\ref{equation:constraint5} is the eigenvalue problem ($EVP$) that could be optimized by Interior-Point Methods \cite{boyd1994linear}.
Through the optimization, we can obtain the optimal numerical solution ${D}$.

\textbf{Modification for computation.} The constraint Eq. \ref{equation:constraint6} limits the boundary of control efficiency. But it is not suitable for practical computation. For convenience of computation, the constraint Eq. \ref{equation:constraint6} is replaced by
\begin{linenomath}
\begin{numcases}
\quad\mathbf{w}^{'}{D}\mathbf{w}<C+\varepsilon,\label{equation:constraint3}\\
C-\varepsilon<\mathbf{w}^{'}D\mathbf{w},\label{equation:constraint4}
\end{numcases}
\end{linenomath}
where $\varepsilon$ is a small positive decimal ($0<\varepsilon\ll C$).
Equation \ref{equation:constraint3}--\ref{equation:constraint4} guarantee that $E_{sum}\rightarrow C$ when $\varepsilon\rightarrow 0$. In the paper, we set $\varepsilon=0.001$.
By synthesizing Eq. \ref{equation:constraint0}--\ref{equation:constraint5} and Eq. \ref{equation:constraint3}--\ref{equation:constraint4}, the optimal feedback gains for all nodes could be obtained under fixed precision.

%

\clearpage

\noindent \textbf{Acknowledgement.} \\
The author thank Dr. Matus Medo for their fruitful discussion and comments. This work is jointly supported by the National Nature Science Foundation of China (Nos. 60974079, 61004102, 61170076 and U1301252),
China Postdoctoral Science Foundation (No. 2013M541840), the Fundamental Research Funds for the Central Universities
(No. ZYGX2012J075) and China Scholarship Council.\\

\noindent \textbf{Author contributions} \\
M.-Y.Z., Z.Z and H.L. designed the research. M.-Y.Z., Z.Z., H.L. and S.-M.C. performed the experiments. S.-M.C. and Z.-Q.F. analyzed the data and improved the method. H.L., Z.-Q.F. and Z.Z. wrote the manuscript. H.L., S.-M.C. and M.-Y.Z. completed the supplementary.

\noindent \textbf{Competing financial interests:}\\
The authors declare no competing financial interests.

\end{document}